\documentclass[12pt]{article}
\usepackage{amsmath}
\usepackage{amssymb}
\usepackage{amsthm}
\usepackage[titletoc,title]{appendix}
\usepackage{graphicx}
\usepackage{float}
\usepackage{wrapfig}
\numberwithin{equation}{section}
\usepackage[left=0.98in,right=0.98in,bottom=1.5in]{geometry}
\usepackage{setspace}
\usepackage [linktocpage]{hyperref}
\begin{document}

\title{\bf Massive (pesudo)Scalars in AdS$_4$, SO(4) Invariant Solutions and Holography \ }

\author{{\bf M. Naghdi \footnote{E-Mail: m.naghdi@ilam.ac.ir} } \\
\textit{Department of Physics, Faculty of Basic Sciences}, \\
\textit{University of Ilam, Ilam, West of Iran.}}
\date{\today}
 \setlength{\topmargin}{0.1in}
 \setlength{\textheight}{9.2in}
  \maketitle
  \vspace{-0.0in}
    \thispagestyle{empty}
 \begin{center}
   \textbf{Abstract}
 \end{center}
We include a new 7-form ansatz in 11-dimensional supergravity over $AdS_4 \times S^7/Z_k$ when the internal space is considered as a $U(1)$ bundle on $CP^3$. After a general analysis of the ansatz, we take a special form of it and obtain a scalar equation from which we focus on a few massive bulk modes that are $SU(4) \times U(1)$ R-singlet and break all supersymmetries. The mass term breaks the scale invariance and so the (perturbative) solutions we obtain are $SO(4)$ invariant in Euclidean $AdS_4$ (or $SO(3,1)$ in its $dS_3$ slicing). The corresponding bare operators are irrelevant in probe approximation; and to realize the AdS$_4$/CFT$_3$ correspondence, we need to swap the fundamental representations of $SO(8)$ for supercharges with those for scalars and fermions. In fact, we have domain-walls arising from (anti)M5-branes wrapping around $S^3/Z_k$ of the internal space with parity breaking scheme. As a result, the duals may be in three-dimensional $U(N)$ or $O(N)$ Chern-Simon models with matters in fundamental representations. Accordingly, we present dual boundary operators and build instanton solutions in a truncated version of the boundary ABJM action; and, because of the unboundedness of bulk potential from below, it is thought that they lead to big crunch singularities in the bulk.

\newpage
\setlength{\topmargin}{-0.7in}
\pagenumbering{arabic} 
\setcounter{page}{2} 

\section{Introduction}
In the past few years, we have studied nonperturbative and localized objects in M2- and D2-brane theories; see \cite{Me6} and \cite{Me5} as our recent studies. The focus was on Instantons, as classical solutions to Euclidean Equations of Motion (EOM) with finite actions, which contribute to phase integrals and mediate tunneling among vacua. We have searched for such objects in both gravity and field theories of AdS/CFT correspondence \cite{Maldacena} as a leading framework to deal with many physical problems. To do this, we have employed the Aharony-Bergman-Jafferis-Maldacena (ABJM) model \cite{ABJM}, as the standard version of AdS$_4$/CFT$_3$ duality, which describes $N$ M2-branes in tip of a $C^4/Z_k$ cone whose near horizon geometry is $AdS_4 \times S^7/Z_k$ with $\acute{N}= k\, N$ units of 4-form flux on $S^7$; and boundary theory is a three-dimensional (3D) $U(N)_k \times U(N)_{-k}$ Chern-Simon (CS) gauge theory with $\mathcal{N}=6$ supersymmetry (SUSY) and  matter fields in bi-fundamental representations (reps) of $SU(4)_R \times U(1)_b$, where $R$ and $b$ indices are for the boundary R- and baryonic-symmetry. For $k=1,2$ the symmetry is enhanced to $SO(8)\equiv G$ and SUSY to $\mathcal{N}=8$  by monopole-operators. In addition, when $k$ becomes large, a suitable description is in terms of type IIA supergravity (SUGRA) over $AdS_4 \times CP^3$ with $S^7$ taken as a $S^1$ fibration on $CP^3$.

Here we include a new 7-form ansatz in the 11D SUGRA background of the ABJM model without changing the geometry. As the main result, we arrive at a second-order NonLinear Partial Differential Equation (NLPDE) for the (pseudo)scalar fields with negative, zero and positive mass squared \footnote{It is notable that the resultant bulk equation is twofold and has solutions corresponding to both Wick-rotated and skew-whiffed versions of the original background.}, which could of course be understood as a consistent truncation in that just the singlet (pseudo)scalars are included in resulting truncated theory; see for instance \cite{LiuSati}. Among the massive modes, which corresponds to irrelevant boundary operators, to be more precise, we focus on the three bulk modes $m^2 R_{AdS}^2 = +4, +10, +18$ that have physical implications such as in (super)conductivity applications of similar arrangements; see \cite{Gauntlett0912} and \cite{Bak1003}.

Because of scale invariance breaking (SIB) by the mass term of the scalar equation in 4D Euclidean Anti-de Sitter space ($EAdS_4$), there is no exact solution with finite action and so, perturbative methods will be main tools to obtain approximate solutions. As a result, we try to solve the bulk equation approximately and in probe approximation, which in turn means ignoring the backreaction of the solutions on the background geometry; see \cite{Me4} for a analogues analysis with massless modes. Doing so, we see that there are $SO(4)$ invariant Euclidean solutions for the massive case while the massless equations have a known $SO(4,1)$ invariant solution of Fubini-type \cite{Fubini1}.

Then, we do an SUGRA mass spectroscopy to see under which conditions our $SU(4) \times U(1) \equiv H$-singlet modes are realized. Based on this, we understand that both swappings $\textbf{8}_s \leftrightarrow \textbf{8}_c$ and $\textbf{8}_s \leftrightarrow \textbf{8}_v$, coming from triality of the prime $G$ symmetry that means exchanging supercharge- with fermion- and scalar- reps respectively, are required to meet the demand that our bulk ansatz breaks all SUSY's as well. In addition, there is the parity-symmetry breaking associated with wrapping the included (anti)M5-branes around $S^3/Z_k$ with an interpretation as fractional (anti)M2-branes \cite{ABJ}. Next, we propose the dual irrelevant operators in leading order \footnote{Note that, because of SIB, the operators gain anomalous dimensions. Here we consider the bare operators $\Delta_\mp$ in leading order, where the scale is nearly invariant.} with which we deform the boundary action, get approximate instanton solutions and confirm the bulk/boundary correspondence according to the well-known AdS/CFT rules. Particularly, there is a special perturbative solution for the massive modes that, with a mixed boundary condition, corresponds to a marginal deformation in leading order.

Further, we will see that the bulk potential has two local maxima and a local minimum and is unbounded from below that in turn signals instability as rolling down from the potential tops or decay from the false vacua. On the other hand, the included (anti)M5-brane has three common directions with the background M2-branes and thus we have a thin- or domain-wall in the bulk that may tunnel between a false vacuum (out of the bubble) and a true vacuum (inside the bubble) or interpolate between two $AdS$ spaces with different radii (AdS$_+$ and AdS$_-$ respectively). Meantime, the domain wall is in a constant $u$ and the breaking of conformal invariance results in renormalization group flows between the fixed-points UV ($u=0$; CFT$_+$) and IR ($u=\infty$; CFT$_-$) of the boundary theory; see \cite{Barbon1003}.

This paper is organized as follows. In section 2, we discuss the (super)gravity side of our study. In particular, in subsection 2.1, we represent the 11D SUGRA background and related conventions; and in subsection 2.2, we introduce a general 7-form ansatz and solve related equations and identities. From the general ansatz, we take a special 4-form in subsection 2.3, obtain the main (pseudo)scalar EOM and discuss a few related issues briefly; while in subsection 2.4, we deal with solutions for the bulk equation. In section 3, we analysis the mass spectrum of the involved SUGRA model and included states. In section 4, we discuss the symmetries of the bulk solutions and conditions that boundary counterparts must obey. With a brief introduction of required AdS$_4$/CFT$_3$ correspondence rules in subsection 5.1, we discuss the boundary field theory duals of the bulk massive modes as pseudoscalars and scalars in subsections 5.2 and 5.3, respectively, including the corresponding operators, solutions and some related discussions. Section 6 is allocated to a brief summary with a few points not addressed in the main text such as interpretations of solutions and the issue of vacuum instability in our setup.

\section{On the Supergravity Side} \label{sub01AA}
In this section, we first present the background 11D SUGRA. Then, we introduce a general 7-form ansatz and drive its corresponding equations and solutions. Next, concentrating on a special 4-form ansatz of the general one, we obtain the main second-order NLPDE in $EAdS_4$ from the 11D equations and identities; and finally try to solve it approximately, ignoring the backreaction, with some solutions in subsection (\ref{sub2.2}.

\subsection{The Background Gravity} \label{sub01}
The background we use is 11D SUGRA over $AdS_4 \times S^7/Z_k$ with metric and 4-form flux as
\begin{equation}\label{eq001}
\begin{split}
& ds^2_{11D}= \frac{R^2}{4} ds^2_{AdS_4} + R^2 ds^2_{S^7/Z_k}, \\
& \ G_4^{(0)}= d\mathcal{A}_3^{(0)} = \frac{3}{8} R^3 \mathcal{E}_4 =  N \mathcal{E}_4,
\end{split}
\end{equation}
where $R=2R_{AdS}=R_7$ is the curvature radius of $AdS$ and $N$ is the 4-flux units on the quotient space. We consider, as ABJM, $S^7/Z_k$ as a $S^1/Z_k$ Hopf-fibration on $CP^3$
\begin{equation} \label{eq001b}
ds_{S^7/Z_k}^2 =ds_{CP^3}^2 + e_7^2, \quad e_7=\frac{1}{k}(d\varphi + k\omega),
\end{equation}
in which $0 \leq \varphi \leq 2\pi$ is the fiber coordinate and $J(=d\omega)$ is the K\"{a}hler form on $CP^3$. We use, in upper-half Poincar$\acute{e}$ coordinate, the Euclidean $AdS_4$ metric and its unit-volume form
\begin{equation}\label{eq001a}
 ds^2_{EAdS_4} = \frac{1}{u^2} \big(du^2 + dx^2 + dy^2 + dz^2 \big),
\end{equation}
\begin{equation}\label{eq001aa}
 \mathcal{E}_4 = \frac{1}{u^4}\ dx \wedge dy \wedge dz \wedge du,
\end{equation}
respectively; while for the 7D unit-volume form we take
\begin{equation}\label{eq001aaa}
 \mathcal{E}_7 = \frac{1}{3!}\ J^3 \wedge e_7.
\end{equation}

\subsection{The 7-Form Ansatz and Equations} \label{sub2.2A}
We introduce the combined 7-from (associated with electric (anti)M5-branes)
\begin{equation}\label{eq11}
 \begin{split}
G_7 = & f_1\, G_4^{(0)} \wedge J \wedge e_7 + df_2 \wedge \mathcal{A}_3^{(0)} \wedge J \wedge e_7 + \ast_4 df_3 \wedge
J^2 + f_4\, \mathcal{A}_3^{(0)} \wedge J^2  \\
& + df_5 \wedge \ast_4 \mathcal{A}_3^{(0)} \wedge J^2 \wedge e_7 + df_6 \wedge J^3 +  f_7\, J^3 \wedge e_7,
 \end{split}
\end{equation}
where $f_1, f_2,\ldots, f_7$ are the functions of $EAdS_4$ coordinates.

From the Bianchi identity $d\ast G_7=0$, where $\ast$ is the 11D Hodge-star, we come to
\begin{equation}\label{eq12d}
    df_4 \wedge \ast_4 \mathcal{A}_3^{(0)}=0,
\end{equation}
\begin{equation}\label{eq12a1}
     c_1\, f_1 + c_2\, d \ast_4(df_2 \wedge \mathcal{A}_3^{(0)})=0,
\end{equation}
\begin{equation}\label{eq12a2}
  c_3\, df_3 + c_4\, f_4 \ast_4 \mathcal{A}_3^{(0)}=0,
\end{equation}
\begin{equation}\label{eq12b}
      c_5\, d \ast_4(df_5 \wedge \ast_4 \mathcal{A}_3^{(0)}) - c_6\, \ast_4 df_6=0,
\end{equation}
\begin{equation}\label{eq12c}
    d(\ast_4 df_6)=0,
\end{equation}
in which the constants $c_1, c_2,\ldots,c_7$ are from
\begin{equation}\label{eq12c1}
 \begin{split}
& \ast \big(G_4^{(0)} \wedge J \wedge e_7 \big)= c_1\, J^2, \quad \ast_7 \big(J \wedge e_7 \big)= c_2\, J^2, \quad \ast \big(\ast_4 df_3 \wedge J^2 \big)= c_3\, df_3 \wedge J \wedge e_7, \\
& \ \ \ \ast_7 J^2= c_4\, J \wedge e_7, \quad \ast_7 \big(J^2 \wedge e_7 \big)= c_5\, J, \quad \ast_7 J^3 = c_6\, e_7, \quad \ast \big(J^3 \wedge e_7 \big)= c_7\, G_4^{(0)},
 \end{split}
\end{equation}
where the Hodge-stars are taken with respect to (wrt) the corresponding metrics in subsection \ref{sub01} and so
\begin{equation}\label{eq12c2}
  c_1= 3, \quad c_2= \frac{R}{2}=-c_3^{-1}= c_4^{-1}, \quad c_5=\frac{2}{R^3}, \quad  c_6=\frac{6}{R^5}, \quad c_7=\frac{1}{R^6},
\end{equation}
where to obtain $c_3$ we have used
\begin{equation}\label{eq12c3}
\varepsilon_{\mu\nu \rho m n p q \sigma r s 7}=+\varepsilon_{\mu \nu \rho \sigma} \varepsilon_{m n p q r s 7}, \quad \ast_4 \ast_4 df_3= -df_3,
\end{equation}
with Greek indices for the external space and Latin ones for the internal $CP^3$ space and 7 for the seventh one.

Then, (\ref{eq12a1}) with $f_1=f_2$ gives
\begin{equation}\label{eq13a1}
  \frac{d^2f_1(u)}{du^2} -\frac{2}{u}\frac{df_1(u)}{du}=0 \Rightarrow f_1(u) = C_{1} + C_{2}\, {u^3},
\end{equation}
with arbitrary constants $C_{1}, C_{2},\ldots$; and from (\ref{eq12a2}) we can write
\begin{equation}\label{eq13a2}
 \ast_4 \mathcal{A}_3^{(0)}=\frac{R}{2u} du \Rightarrow f_4(u)=\frac{2}{R} u \frac{df_3(u)}{du},
\end{equation}
and then (\ref{eq12d}) satisfies trivially; while $(\ref{eq12b})$ satisfies with
\begin{equation}\label{eq13a3}
 f_5=f_5(x,y,z), \quad f_6=-\frac{2}{3} R f_5,
\end{equation}
and so from (\ref{eq12c}) we read
\begin{equation}\label{eq13a4}
 \partial_i \partial^i f_6 = 0; \quad f_5(r)=C_{3} + \frac{C_{4}}{r},
\end{equation}
where $r=\sqrt{x_i x^i}$ with $i=1,2,3$ of $x_i$ for $x,y,z$ respectively.

We should also survey the conditions to satisfy the Euclidean EOM
 \begin{equation}\label{eq14a0}
  d \ast G_4 - \frac{i}{2} G_4 \wedge G_4=0,
\end{equation}
 with the anstaz (\ref{eq11}). As a result, from the terms including $J^2$, we should take $f_2=f_2(u)$, which could be as (\ref{eq13a1}), in addition to the condition (\ref{eq12a2}) (for $f_3$ and $f_4$) and that
\begin{equation}\label{eq14a}
 f_7= - \frac{i}{3} R^3.
\end{equation}
As the same way, the terms including $J^3$ are satisfied with $f_5$ in (\ref{eq13a3}) and $f_1$ in (\ref{eq13a1}) with
$C_{1}=0$. Similarly, the terms including $J^2 \wedge e_7$ are satisfied 
with the latter plus (\ref{eq13a2}), and those including $J^3 \wedge e_7$ are satisfied with the last condition.

\subsection{The 4-Form Ansatz, Equation and Bulk Modes} \label{sub3.1A}
An interesting case is when we consider the first, third and seventh terms of the main ansatz (\ref{eq11}). Therefore, for its 11D dual, we write
\begin{equation}\label{eq18}
\tilde{G}_4 = \frac{8}{R^3} R^{d_1} \bar{f}_1 J^2 - \frac{2}{R} R^{d_3} df_3 \wedge J \wedge e_7 + \frac{3}{8 R^3}
R^{d_7} f_7 \mathcal{E}_4,
\end{equation}
in which we have redefined $f_1 N =\bar{f_1}$ for convenience and $R^{d_1}$, $R^{d_3}$ and $R^{d_7}$ are introduced as the dimensional coefficients. Now, from the Bianchi identity $d\tilde{G}_4=0$ we have
\begin{equation}\label{eq19}
 df_3= - \frac{4}{R^2} \frac{R^{d_1}}{R^{d_3}} d\bar{f}_1 \Rightarrow \bar{f}_1 = - \frac{R^2}{4} \frac{R^{d_3}}{R^{d_1}}
 f_3 \pm \bar{c}_1,
\end{equation}
with $\bar{c}, \bar{C}, \hat{c}, \hat{C}$'s (such as $c, C$), with lower indices, as integration constants throughout. From the EOM for $\tilde{G}_4$, one condition with help of (\ref{eq19}) reads
\begin{equation}\label{eq20a}
 df_7 = i \frac{16 \times 4}{R^6} \frac{R^{2d_1}}{R^{d_7}} \bar{f}_1 d\bar{f}_1 \Rightarrow f_7 = +i \frac{32}{R^6} \frac{R^{2d_1}}{R^{d_7}} \bar{f}_1^2 \pm i \bar{c}_7,
\end{equation}
and inserting it into another condition results in
\begin{equation}\label{eq20b}
     \ast_4 d \left(\ast_4 d\bar{f}_1 \right) - \frac{4}{R^2} \bar{f}_1 \mp  \frac{12}{R^8} R^{d_7} \bar{c}_7 \bar{f}_1 -
     \frac{2\times 192}{R^{14}} R^{2d_1} \bar{f}_1^3=0.
\end{equation}
Next, from the dimensional analysis, we set $d_1=d_7=7$ and $\bar{c}_7 = \frac{\bar{C}_7}{R}$, and take $\lambda_4=192$. Therefore, for various $\bar{C}_7$'s, we have towers of massive and tachyonic bulk (pseudo)scalars besides the massless one with $\bar{C}_7 =\frac{1}{3}$ for the skew-whiffed case. In particular, we have the conformally coupled pseudoscalar ($m^2 R_{AdS}^2=-2$) for $\bar{C}_7=+1$ (the skew-whiffed background) and the non-minimally coupled one ($m^2 R_{AdS}^2=+4$) for $\bar{C}_7=-1$ (the Wick-rotated background) as in \cite{Me6} except for the coupling $\lambda_4=192$ here versus $\lambda_4=3$ there \footnote{It is notable that the equation (\ref{eq20b}) is similar in nature to a consistent truncation of M-theory on $AdS_4 \times S^7$ \cite{Duff99}, from which the $\mathcal{N}=8$ $SO(8)$ gauged supergravity in four dimensions is obtained. By restricting to the Cartan subgroup $U(1)^4$ of the latter group, the resultant Lagrangian includes scalars (dilatons and axions), gauge fields and graviton. Still, in a special case, one may keep just graviton next to a scalar with the Lagrangian
\begin{equation}\label{eq20bb}
   \begin{split}
    & \ \ \ \ \ \ \ \ \ \ \ \ \ \ L_4^E \approx - \mathcal{R}_4 +\frac{1}{2}({\partial}_{\mu} \varphi)({\partial}^{\mu}\varphi) + V(\varphi), \\ & V(\varphi)= - \frac{4}{R^2}\big(2+\cosh(\sqrt{2} \varphi)\big) = - \frac{1}{R^2}\big(12 + 4 \varphi^2 +\frac{2}{3} \varphi^4 + ...\big).
   \end{split}
\end{equation}
We note that the first term on RHS of the potential is the $AdS$ vacuum solution (the cosmological constant $\Lambda$) that happens for $\varphi=0$; and small fluctuations around it have the mass $m^2 R_{AdS}^2=-2$ (the second term).}.

In this study, besides a brief discussion on the conformal mode $m^2 R_{AdS}^2=-2$ and new discussions on the mode $m^2 R_{AdS}^2=+4$, we deal with the new modes $m^2 R_{AdS}^2=+10$ and $m^2 R_{AdS}^2=+18$, where the latter two realize with $\bar{C}_7= 3$ and $\bar{C}_7=\frac{17}{3}$ in the Wick-rotated case of (\ref{eq20b}), respectively. Note also that the last three modes are coupled non-minimally with gravity and that although we have seen the appearance of the first three modes in the preceding discussions, the second and last modes are indeed recognized with the \emph{squashing} and \emph{breathing} modes in \cite{Gauntlett03}, and are suitable for bulk/boundary considerations here as well.

\subsection{Solutions For the Scalar Equation} \label{sub2.2}
To find solutions with the bulk modes we consider, we write the equation (\ref{eq20b}) with $\bar{f}_1\equiv f$, ignoring backreaction, as follows
\begin{equation}\label{eq34}
     \left(\frac{\partial^2}{\partial r^2} + \frac{2}{r} \frac{\partial}{\partial r} \right) f(u,r) + \left(\frac{\partial^2}{\partial u^2} - \frac{2}{u} \frac{\partial}{\partial u} \right) f(u,r) - \frac{1}{u^2} \left[ \hat{C}_1\, f(u,r)+ \hat{C}_2\, f(u,r)^3 \right]=0,
\end{equation}
where
\begin{equation}\label{eq34a}
   \hat{C}_1=(1 \pm 3 \bar{C}_7)={m^2} {R_{AdS}^2}, \quad \hat{C}_2 = 2\, \lambda_4 R_{AdS}^2.
\end{equation}
This second-order NLPDE is of Elliptic type and seems that it does not have any closed solution (except the trivial one $f= \pm\, i \sqrt{\frac{\hat{C}_1}{\hat{C}_2}}$) at least because of the mass term that breaks the scale invariance (SI). Searching for solutions in general is outside the aim of this study. Nevertheless, with specific methods to solve this type of equations, here we discuss a few solutions suitable for our considerations with emphasis on the case $\hat{C}_1 =4$ in that the method is same for $\hat{C}_1 =10, 18$.

By discarding the nonlinear term in the equation for now and employing the method of \emph{separation of variables} as $f(u,r)=f(r)\, g(u)$, one can easily obtain a solution with \emph{Hyperbolic}- and \emph{Bessel}- functions. In the simplest case, the solution reads
\begin{equation}\label{eq35}
   f(r)= C_{5} + \frac{C_{6}}{r}, \quad g(u)= C_{7}\, u^{\Delta_-} + C_{8}\, u^{\Delta_+}, \quad \Delta_{\mp}=\frac{3}{2} \mp \frac{1}{2} \sqrt{9 + 4\, \hat{C}_1};
\end{equation}
and we note that to have real $\Delta_{\mp}$'s for negative $\hat{C}_1$'s, we must take $\hat{C}_1 \leq \frac{9}{4}$, which is nothing but the Breitenlohner-Freedman (BF) bound \cite{BreitenlohnerFreedman} to which we return in section \ref{sec.6}. By inserting the linear solution into the main equation (\ref{eq34}), there is not any nontrivial case to be considered. However, putting the solution (\ref{eq35}) instead of the function in the nonlinear term of (\ref{eq34}), one can obtain a perturbative solution and, after a series expansion about $u$, keep the terms suitable for the boundary studies. In general, by using various ansatzs and methods, one may get some perturbative solutions whose series expansion around $u$, with keeping just the terms appropriate for the boundary studies of the associated operator, can be written as
\begin{equation}\label{eq36AA}
   f(u,r)= f_1(u,r)\, u^{\Delta_-} + f_2(u,r)\, u^{\Delta_+},
\end{equation}
where $f_1(u,r)$ and $f_2(u,r)$ are polynomial, trigonometric, hyperbolic and logarithm functions of $r, u$ depending on the method used.

Still, as an example and in order to perform the boundary calculations, we use the ansatz
\begin{equation}\label{eq38a}
   f(u,r)=F(\xi), \quad \xi=u^{3/2}\ f(r)
\end{equation}
that results in the nonlinear ordinary differential equation
\begin{equation}\label{eq38b}
\left(\frac{d^2}{d \xi^{2}}-\frac{1}{\xi}
\frac{d}{d\xi} \right)F \left( \xi \right)-\frac{4}{9 {\xi}^{2}} \left[ \hat{C}_1\,F \left( \xi \right)+  \hat{C}_2\, F \left( \xi \right)^{3}\right]=0,
\end{equation}
whose linear-part solution (valid for $u\rightarrow 0$ too) is alike $g(u)$ in (\ref{eq35}) with $u\rightarrow u f(r)$. By substituting the latter solution into the nonlinear part of the equation, two proper terms of the first order perturbative solution read
\begin{equation}\label{eq38CC}
    f(u,r)= {C_{9}\, \left(u\, f(r)^{2/3}\right)^{\Delta_-}} + {C_{10}}\, \left(u\, f(r)^{2/3} \right)^{\Delta_+}.
\end{equation}

Alternatively, we may write our $EAdS_4$ Laplacian as
\begin{equation}\label{eq42}
 \Box_4\, f = -\frac{16\, u}{R^3}\, g + \frac{8\, u^3}{R^3}\, \left(\partial_i \partial_i + \partial_u \partial_u \right), \quad f=\frac{2\, u}{R}\, g,
\end{equation}
and thus the main equation (\ref{eq34}) becomes
\begin{equation}\label{eq42a}
\left(\partial_i \partial_i + \partial_u \partial_u \right)\, g -\frac{(2+\hat{C}_1)}{u^2}\, g - \hat{C}_2\, g^3 =0.
\end{equation}
On the other hand, we note that there is an exact solution with $\hat{C}_1=-2$, the conformally coupled (pseudo)scalar, as follows
\begin{equation}\label{eq43a}
g_0(u,\vec{u}) =\sqrt{\frac{2}{\hat{C}_2}}\, \frac{2\, b_0}{-b_0^2 + (a_0+u)^2 + (\vec{u}-\vec{u}_0)^2},
\end{equation}
where $\vec{u}=(x,y,z)$; see \cite{Me6}. To get a rough solution, we use $g_0$ as the initial data and the term including $u$ in the equation (\ref{eq42a}) as a perturbation. As a result, we arrive at a solution whose series expansion up to the first order, for general $\hat{C}_1$'s (except $\hat{C}_1=-2$) reads
\begin{equation}\label{eq43c}
f^{(1)}(u,r) = \sqrt{\frac{2}{\hat{C}_2}}\, {\frac{8\,a_0\, b_0}{3\, R}}\, \frac{(\hat{C}_1+14)(a_0^2+b_0^2 + r^2)}{(a_0^2-b_0^2 + r^2)^4} u^4 + O(u, u^2,u^3,...),
\end{equation}
with a similar structure for the terms including $u^5$ and $u^6$. When discussing dual symmetries and boundary solutions, we see that the vacuum-expectation-values (vev's) of the proposed operators match with these solutions. Nevertheless, we notice that because of the mass term in (\ref{eq34}), the SI is broken and so, the massless solution (\ref{eq43a}) and other similar ones are valid approximately. It is also notable that there is a well-known estimated solution, valid for $m\, b_0 \ll 1$, introduced in \cite{Affleck1} as \emph{constrained instanton}.

\section{The Bulk Mass-Spectrum} \label{sub.1.4}
From compactification of 11D SUGRA on $S^7$, an effective 4D theory for $AdS_4$ is obtained, which includes an infinite tower of massless and massive states with the masses quantized through $m$ (proportional to the inverse radius of $S^7$). These states are classified into multiplets of $Osp(8\mid 4)$ ($\supset SO(3,2) \times SO(8)$) with the maximum spin 2 for a multiplet ($s\leq 2$) and that the unitarity of a rep is provided that $E_0 = \Delta \geq s+\frac{1}{2}$ for $s=0, \frac{1}{2}$, where $E_0$ is the lowest eigenvalue of the energy operator of the subalgebra $SO(3,2)$; For earlier studies in the case see, for instances, \cite{Biran}, \cite{FreedmanNicolai}, \cite{DuffNilssonPope84} and \cite{GünaydinWarner}.

The massless modes \footnote{The multiplet is massless in a sense that the masses of scalars are shifted by $-\mathcal{R}_4/{6}$ and so $\tilde{m}^2 R_{AdS}^2 ={m}^2 R_{AdS}^2 +2$, where $\tilde{m}$ is the mass that appears in supergravity literatures.} on $S^7$ include a graviton ($\textbf{1}(0,0,0,0)$), a gravitino ($\textbf{8}_s(0,0,0,1)$), 28 spin-1 fields ($\textbf{28}(0,1,0,0)$), 56 spin-$\frac{1}{2}$ fields ($\textbf{56}_{s}(1,0,1,0)$), 35 scalars ($\textbf{35}_{v}(2,0,0,0)$) with $\Delta_- =1$ and 35-pseudoscalars ($\textbf{35}_{c}(0,0,2,0)$) with $\Delta_+ =2$. Because of the positive energy theorems of SUGRA \cite{BreitenlohnerFreedman}, the fields should be in Unitary Irreducible Representations (UIR's) of $Osp(8\mid 4)$. The massive UIR's, which may be reducible under $G$, are obtained from the tensor products of the massless multiplet and a representation with the Dynkin labels $(n,0,0,0)$, which in turn corresponds to eigenmodes of the scalar Laplacian on $S^7$ and to symmetric and traceless tensors of $G$ with $n$ indices, and $n\in N$ labels massive levels,
\begin{equation}\label{eq47}
      (n,0,0,0) \otimes \{ \textbf{1}, \textbf{8}_s, \textbf{28}, \textbf{56}_{s}, \textbf{35}_{v}, \textbf{35}_{c} \}.
\end{equation}
After the Hopf reduction ${S^7}/{Z_k} \approx CP^3 \ltimes S^1$ that we consider, only neutral states under $U(1) \sim SO(2)$ remain in the spectrum \cite{NilssonPope}. In other words, in the large $k$ limit, only the singlet $U(1)$ states remain and the states with odd $n$ on $S^7$ are excluded in that they lead to charged $U(1)$ states \cite{Bianchi2}.

Then, on the one hand, we note that the ansatz (\ref{eq18}) is $H$-singlet in that $e_7$ and $J$ are so. On the other hand, we know that there are three generations of scalars ($0^+_1, 0^+_2, 0^+_3$) and two of pseudoscalars ($0^-_1, 0^-_2$) in the spectrum; see for instance \cite{FreedmanNicolai}. Now, for three (pseudo)scalars $m^2 R_{AdS}^2 = 4, 10, 18$ that we consider, the only singlets under the branching $G\rightarrow H$ in the original background appear as
\begin{equation}\label{eq48a}
      \textbf{35}_{s}(0,0,0,2) \rightarrow \textbf{1}_{-4} \oplus \textbf{1}_{0}[0, 0, 0] \oplus {\textbf{1}}_{4} \oplus {\textbf{6}}_{-2}[0, 1, 0] \oplus \textbf{6}_{2} \oplus \acute{\textbf{20}}_{0}[0, 2, 0],
\end{equation}
in $0^-_2$ for $\Delta_+ =5$ with $n=2$ and
\begin{equation}\label{eq48b}
 \begin{split}
  \textbf{300}(0,2,0,0) \rightarrow &  \textbf{1}_{0}[0, 0, 0] \oplus {\textbf{6}}_{-2}[0, 1, 0] \oplus \textbf{6}_{2} \oplus {\textbf{15}}_{0}[1, 0, 1] \oplus \acute{\textbf{20}}_{-4}[0, 2, 0] \oplus \acute{\textbf{20}}_{0} \oplus \acute{\textbf{20}}_{4} \\
  & \oplus {\textbf{64}}_{-2}[1, 1, 1] \oplus \textbf{64}_{2} \oplus \textbf{84}_{0}[2, 0, 2],
 \end{split}
\end{equation}
in $0^+_3$ for $\Delta_+ =4$ with $n=2$ and also
\begin{equation}\label{eq48c}
      \textbf{1}(0,0,0,0) \rightarrow \textbf{1}_{0}[0, 0, 0],
\end{equation}
in $0^+_2$ for $\Delta_+ =6$ with $n=2$, respectively, given that $m^2 R_{AdS_4}^2 = \Delta (\Delta -3)$.

Meanwhile, the reps $\textbf{5775}_{vc}(4,0,2,0)$ of $0^-_1$ with $n=4$, $\textbf{24024}_{vc}(6,0,2,0)$ of $0^-_1$ with $n=6$ and $\textbf{75075}_{vc}(8,0,2,0)$ of $0^-_1$ with $n=6$ of $G$ lead to $U(1)$-neutral and non-singlets of $SU(4)$ for $m^2 R_{AdS}^2 = 4, 10, 18$ respectively; while for the latter mode, $\acute{\textbf{840}}_{c}(2,0,0,2)$ of $0^-_2$ with $n=4$ of $G$ has the same behaviour. As scalars, these modes in turn appear in $\textbf{4719}_{v}(8,0,0,0)$ of $0^+_1$ with $n=6$ and $\textbf{300}(0,2,0,0)$ of $0^+_3$ with $n=2$; $\textbf{13013}_{v}(10,0,0,0)$ of $0^+_1$ with $n=8$ and $\textbf{4312}_{v}(2,2,0,0)$ of $0^+_3$ with $n=4$; and $\textbf{30940}_{v}(12,0,0,0)$ of $0^+_1$ with $n=10$ and $\textbf{1}(0,0,0,0)$ of $0^+_2$ with $n=2$ and also $\textbf{23400}_{v}(4,2,0,0)$ of $0^+_3$ with $n=6$ of $G$, respectively; and go to non-singlets of $SU(4)$ under the branching except for the $0^+_2$ mode.

On the other hand, the triality property of $SO(8)$ implies that there are three inequivalent reps $\textbf{8}_v(1,0,0,0)$, $\textbf{8}_s(0,0,0,1)$ and $\textbf{8}_c(0,0,1,0)$, where the $\textbf{s}$- and $\textbf{c}$-type reps occur just for the spins $s=0^-, \frac{1}{2}, \frac{3}{2}$. We make use of this triality to go from the original (left-handed) to the skew-whiffed (right-handed) version of the ABJM model to see whether we find our needed singlet modes or not. Doing so, we see that after exchanging $\textbf{s} \leftrightarrow \textbf{c}$ with $\textbf{v}$ fixed that means exchanging the spinors and fermions while keeping the scalars fixed, the pseudoscalar reps change correspondingly without any $SU(4)$-singlet under the branching while the scalar reps do not change. It is notable that the only singlet pseudoscalar in the original spectrum of our modes, which appeared in $\textbf{35}_{s}$, now appears in
\begin{equation}\label{eq48d}
      \textbf{35}_{c}(0,0,2,0) \rightarrow \textbf{10}_{-2}[2, 0, 0] \oplus \bar{\textbf{10}}_{2}[0, 0, 2] \oplus {\textbf{15}}_{0}[1, 0, 1],
\end{equation}
which does not include any singlet under $H$. In the same way, after exchanging $\textbf{s} \leftrightarrow \textbf{v}$ with $\textbf{c}$ fixed that means exchanging the spinors and scalars while keeping the fermions fixed, we have the scalar reps $\textbf{5775}_{sc}(0,2,0,4)$, $\textbf{24024}_{sc}(0,0,2,6)$, $\textbf{75075}_{sc}(0,0,2,8)$ from the pseudoscalar ones and $\acute{\textbf{840}}_{c}(2,0,0,2)$ unchanged, while $\textbf{35}_{s}$ of (\ref{eq48a}) changes into
\begin{equation}\label{eq48d}
      \textbf{35}_{v}(2,0,0,0) \rightarrow \textbf{10}_{2}[2, 0, 0] \oplus \bar{\textbf{10}}_{-2}[0, 0, 2] \oplus {\textbf{15}}_{0}[1, 0, 1],
\end{equation}
where no singlet under the branching appears again.

In particular, our modes as scalars with the latter swapping and after the branching read
\begin{equation}\label{eq49a}
   \begin{split}
      \textbf{30940}_{s}(0,0,0,12) \rightarrow  & \textbf{1}_{0}[0, 0, 0] \oplus \acute{\textbf{20}}_{0}[0, 2, 0] \oplus \textbf{105}_{0}[0, 4, 0] \oplus \textbf{336}_{0}[0, 6, 0] \\
      & \oplus \acute{\textbf{825}}_{0}[0, 8, 0] \oplus \acute{\textbf{1716}}_{0}[0, 10, 0] \oplus \textbf{3185}_{0}[0, 12, 0] \oplus ...\ ,
   \end{split}
\end{equation}
and same for $\textbf{13013}_{s}(0,0,0,10)$ and $\textbf{4719}_{s}(0,0,0,8)$ except the last and last two terms on its RHS respectively, and
\begin{equation}\label{eq49b}
   \begin{split}
      \textbf{4312}_{s}(2,2,0,0) \rightarrow  & \textbf{1}_{0}[0, 0, 0] \oplus \textbf{15}_{0}[1, 0, 1] \oplus 3\, (\acute{\textbf{20}}_{0})[0, 2, 0] \oplus \textbf{84}_{0}[2, 0, 2] \oplus \textbf{105}_{0}[0, 4, 0] \\
      & 2\,({\textbf{175}}_{0})[1, 2, 1] \oplus {\textbf{729}}_{0}[2, 2, 2] \oplus ...\ ,
   \end{split}
\end{equation}
\begin{equation}\label{eq49c}
   \begin{split}
      \textbf{23400}_{s}(0,2,0,4) \rightarrow  & \textbf{1}_{0}[0, 0, 0] \oplus \textbf{15}_{0}[1, 0, 1] \oplus 3\, (\acute{\textbf{20}}_{0})[0, 2, 0] \oplus \textbf{84}_{0}[2, 0, 2] \oplus 3\, (\textbf{105}_{0})[0, 4, 0] \\
      &\oplus 2\,({\textbf{175}}_{0})[1, 2, 1] \oplus {\textbf{336}}_{0}[0, 6, 0] \oplus {\textbf{729}}_{0}[2, 2, 2] \oplus 2\,({\textbf{735}}_{0})[1, 4, 1] \\
      & \oplus {\textbf{3640}}_{0}[2, 4, 2]  \oplus ...\ ,
   \end{split}
\end{equation}
where we have just kept the neutral $U(1)$ reps. It is also notable that the reps $\textbf{300}(0,2,0,0)$ and $\textbf{1}(0,0,0,0)$ will remain the same as in (\ref{eq48b}) and (\ref{eq48c}), respectively. Therefore, we see that after the swapping $\textbf{8}_s \leftrightarrow \textbf{8}_v$, the $H$-singlets appear in all generations for the three scalar modes we are considering.

\section{Dual Symmetries} \label{sec.5.1}
In the previous section \ref{sub.1.4}, we discussed corresponding states for the (pseudo)scalars in the ansatz (\ref{eq18}), which is $H$-singlet in that $e_7$ and $J$ are so. When searching for dual field theory solutions, we see how to adjust these bulk states to the boundary operators.

On the other hand, we simply read from the ansatz (\ref{eq18}) that the whole supersymmetries are broken as the associated (anti)M-branes wrap around the mixed internal and external directions; in addition to the fact that the solutions with $G_{mnpq} \neq 0$ break SUSY's and parity as well \cite{DuffNilssonPope84}. As a result, we can say that we are indeed adding M-branes or anti-M-branes to the (Wick-rotated)background M2-branes along with breaking all SUSY's and external space isometries while preserving the internal space isometries.

More precisely, we know that the isometry group of $EAdS_4$ is $SO(4,1)$ (or $SO(3,2)$ with Lorentzian signature), which is in turn the conformal symmetry of the boundary CFT$_3$. There are 10 group parameters that include three translations ($P_\mu$), three Lorentz rotations ($L_{\mu \nu}$), one dilation ($D$) and three special conformal transformations ($K_\mu$). In addition, there are five conformal killing vectors because of the conformal flatness of $AdS_4$. The issue now is which symmetries are broken in our setup of the ansatz, equations and solutions. In particular, we find the solutions invariant under the largest subgroup of the main isometry group. To this end, we first pay attention to the SIB, through the mass term in the action from which the equation (\ref{eq20b}) arises, as
 \begin{equation}\label{eq51}
      \bar{f}_1(x) \rightarrow a \bar{f}_1(a x) \Rightarrow m^2 \int d^4x\,  \bar{f}_1^2 \rightarrow m^2 a^{-2}  \int d^4x\,  \bar{f}_1^2,
\end{equation}
where $a$ is the scale parameter. As a result, the tree dimensions ($\Delta_\pm$) of the boundary operators will change and so we consider a rough SI just in leading order.

Second, we note that although the Laplacian ($\square_{4}$) preserves all $AdS$ isometries- in particular it is invariant under translations and the inversion $I: x_\mu \rightarrow \frac{x_\mu}{u^2+r^2}$- but our ansatz breaks the inversion symmetry and so breaks the special conformal transformations $K_\mu$ in that one can write $K_\mu = I P_\mu I$. Further, by having a non-constant solution, the translational invariance breaks as well. Meanwhile, it should be noted that our (perturbative) bulk solutions preserve the Lorentz invariance because of their dependence on $r$.

On the other hand, it is discussed in \cite{Fubini1}, see also \cite{Loran2}, that for the massless (pseudo)scalar of the so-called $\phi^4$ model in Euclidean $AdS_4$ (see (\ref{eq20b}) without the mass term or for the conformally coupled case $m^2 R_{AdS}^2=-2$) a nontrivial solution has $SO(4,1)$ symmetry. The mass term breaks the SI and thus just $SO(4)$ part of the symmetry remains, which in turn becomes the symmetry $SO(3,1)$ of the de sitter space-time $dS_3$ after Lorentz continuation \footnote{Note that the $AdS_4$ boundary is a copy of $R^3$ at $u=0$ together with a point at $u=\infty$, and this is $S^3$, which is the most natural boundary. Meantime, with constant $u$'s, the $dS_3$ slices of $AdS_4$ are realized.}. In fact, the latter group of symmetries has six parameters and is consists of Lorentz transformations $L_{\mu \nu}$ and $L_{\mu 4}\equiv R_\mu \approx (K_\mu + a^2 P_\mu )$, which correspond to rotations on $S^3$-- note that for the bulk massless solution (\ref{eq43a}), $S_\mu \approx (K_\mu - b_0^2\, P_\mu)$ is used in place of $R_\mu$. Thus, wrt the discussion on the preceding paragraph, we will see that our boundary solutions respect the latter symmetry that means the breaking of conformal symmetry as well.

As another aspect, the main ABJM model has even parity that means interchanging the levels ($k\rightarrow -k$) of the quiver gauge group $SU(N)_k \times SU(M)_{-k}$. The breaking of parity invariance in our setup is related to the idea of fractional M2-branes \cite{ABJ}, as M5-branes wrapping around three internal directions (${S^3}/{Z_k}$), reading from the first term of the ansatz (\ref{eq11}). Because of the parity and supersymmetry breaking, the boundary theories may be Chern-Simon-matter $U(N)$ and $O(N)$ vector models, which are in turn dual to the bulk Vasiliev's Higher-Spin theories \cite{Vasiliev01}; see \cite{Giombi01} and references therein as a useful review. We return to these issues when discussing the dual field theory solutions.

\section{On the Field Theory Duals} \label{sec.6}
In this section, we first present a brief discussion of the AdS$_4/$CFT$_3$ rules, and then analysis the field theory duals for the bulk massive modes we are considering as pseudoscalars and scalars. It is noticeable that the discussions here are corresponding to the bulk solutions, as discussed in subsections \ref{sub3.1A} and \ref{sub2.2}, without including the backreaction.

\subsection{Basic Correspondence}
First, we note that solutions to the wave equation (\ref{eq34}), near the boundary ($u=0$), have a series expansion as
\begin{equation}\label{eq52}
      {f}(u,\vec{u}) \approx  u^{\Delta_-} (\alpha(\vec{u}) + \ldots) + u^{\Delta_+} (\beta(\vec{u}) + \ldots),
\end{equation}
where $\alpha$ and $\beta$ act as source and vev for the operator $\Delta_+$ and conversely for $\Delta_-$. Second, we recall that for the (pseudo)scalar fields with $-\frac{9}{4} \leq m^2 R_{AdS}^2 \leq -\frac{5}{4}$ in $AdS_4$, there are two consistent ways to impose boundary conditions; while with larger masses there is one way to ensure the normalizability of perturbations. In fact, with $m^2 R_{AdS}^2 > -\frac{5}{4}$, the only normalizable mode is for $\Delta_+$. Meanwhile, to ensure the reality of the conformal dimensions, there is the so-called BF bound $m^2 R_{AdS}^2 \geq -\frac{9}{4}$ \cite{BreitenlohnerFreedman} and the (pseudo)scalars satisfying this bound are stable naively. Therefore, for our bulk modes $m^2 R_{AdS}^2 = 4, 10, 18$, the normalizable modes corresponding to Dirichlet boundary condition are permissable; although the scalars theories coupled to gravity in general allow a large class of boundary conditions--note also that the massive bulk fields are dual to single-trace operators, which are of course not conserved currents. Third, although the SI is broken by the mass term in the bulk Lagrangian, here we use the bare dimensions of operators as the leading order approximation used to solve the bulk equation (\ref{eq34}) too.

Fourth, according to the mass spectroscopy presented in section \ref{sub.1.4}, to build the boundary operators, we start from
\begin{equation}\label{eq53}
      O^{k} \equiv \texttt{tr} \big(X^{[I_1} X_{[I_2}^\dagger \ldots X^{I_{k-1}]} X_{I_k]}^\dagger \big),
\end{equation}
as the lowest state of the $k$-th Kaluza-Klein supermultiplet or (1/2 BPS) chiral primary operator composed of symmetrized and traceless (\texttt{tr}) products of the (pseudo)scalar fields, where $X^I$'s are 8 free scalars with $SO(8)$ indices. Other states, as descendants, are obtained by applying SUSY transformations, which are $\{Q, X\}\approx \psi$, $\{Q, \psi\}\approx D X$, symbolically; We return to this in next subsections.

Finally, from the well-known AdS/CFT dictionary \cite{KlebanovWitten}, in Euclidean space, we can write
\begin{equation}\label{eq53a}
   \begin{split}
  & \ \ \ \ \ \ S_{on}[\alpha] = -W[\alpha], \quad \langle \mathcal{{O}}_{\Delta_+} \rangle_{\alpha} = - \frac{\delta W[\alpha(\vec{u})]}{\delta\alpha(\vec{u})} = \frac{1}{3} \beta(\vec{u}) \equiv \sigma(\vec{u}), \\
   \alpha(\vec{u}) = & - \frac{\delta W[\sigma(\vec{u})]}{\delta\sigma(\vec{u})}= - \langle \mathcal{{O}}_{\Delta_-} \rangle_{\beta}, \quad \Gamma_{eff.} [\sigma] = - W[\alpha] - \int d^3 \vec{u}\ \alpha(\vec{u})\, \sigma(\vec{u})= W[\sigma],
   \end{split}
\end{equation}
where $S_{on}$ is the on-shell action on $AdS_4$ and $W[\alpha]$ is the generating functional of the connected correlators of CFT$_3$ in which the $O_{\Delta_+}$ dynamics is encoded. $\Gamma_{eff.} [\sigma]$, the Legendre transform of $W[\alpha]$, is the effective action of the local operator $O_{\Delta_+}$ and generating functional ($W[\sigma]$) of the dual theory with ${\Delta_-}$ as well. 

\subsection{Duals For Massive Modes as Pseudoscalars}
With the ansatz (\ref{eq11}), we suspect that our modes are pseudoscalars emerging from the form field (as $\mathcal{A}_{m n p}$) in terms of the internal ingredients; see \cite{NilssonPope}. As the first family of pseudoscalars ($0_1^-$) with Dynkin labels $(n,0,2,0)$ with $n=4,6,8$ for $m^2 R_{AdS}^2 = 4, 10, 18$ respectively, the free field operators read
 \begin{equation}\label{eq54}
      O_{\Delta_+=\frac{n}{2}+2} = \texttt{tr} \big(\Psi^{[I} \Psi^{J^\dagger]}\big)\, O^{n},
\end{equation}
which are indeed the second descendant of those in (\ref{eq53}) with $k=n+2$; see \cite{EDHoker02}. On the way of (\ref{eq54}), we have already formed the $\Delta_+=4$ operator and a solution in \cite{Me6}. One can do similar procedure to build the solutions for $\Delta_+=5, 6$ according to the plain operators
 \begin{equation}\label{eq54a}
     \mathcal{O}_{\Delta_+} = \texttt{tr} \left(\psi_A \psi^{A\dagger} Y^{B_1} Y_{B_1}^\dagger \ldots Y^{B_{\frac{n}{2}}} Y_{B_{\frac{n}{2}}}^\dagger \right),
\end{equation}
where $X^I \rightarrow (Y^A, Y_A^\dagger)$ and $\Psi^I \rightarrow (\psi_A, \psi^{A\dagger})$ with $I,J \ldots =(1,\ldots 6,7,8)=(m,7,8)$, $A=1,2,3,4$, which are realized as $\textbf{8}_v \rightarrow \textbf{4}_{+1} \oplus \bar{\textbf{4}}_{-1}$ and $\textbf{8}_c \rightarrow \textbf{4}_{-1} \oplus \bar{\textbf{4}}_{+1}$ in the original theory, respectively. Still, we remember from section \ref{sub.1.4} that no $H$-singlet rep appears for these modes even after the swapping $\textbf{s} \leftrightarrow \textbf{c}$. Nevertheless, with a limited number of fields and appropriate ansatzs, we may be able to build the desired singlet operators.

Besides, we have $m^2 R_{AdS}^2 = 10$ in the second family of pseudoscalars ($0_2^-$) with Dynkin labels $(0,0,0,2)$ and so, as the sixth descendant of $O^{k}$ with $k=n-2=4$, we arrive at an operator with $\Delta_+=5$ for that we employ the plain form
 \begin{equation}\label{eq55}
      \mathcal{O}_{\Delta_+=5} = \texttt{tr}((\partial X)^2 \Psi \bar{\Psi}),
\end{equation}
which has the wished $H$-singlet according to (\ref{eq48a}). To make a dual solution with the help of the latter operator, note that under the swapping $\textbf{8}_c \leftrightarrow \textbf{8}_s$, where $\textbf{8}_s \rightarrow \textbf{1}_{-2} \oplus \textbf{1}_{+2} \oplus \textbf{6}_{0}$, we can write $\Psi^I \rightarrow (\psi^m, \psi, \bar{\psi})$ with $\psi = \psi^7 + i \psi^8$, $\bar{\psi}= \psi^\dagger$.

On the other hand, although the original ABJM model has even parity, to make a dual for the non-minimal bulk pseudoscalar with parity breaking scheme, we keep just one part of the quiver gauge group next to its CS term together with massless matter fields. This aim is realized through fractional M2-branes idea \cite{ABJ} and the Novel Higgs-mechanism \cite{ChuNastaseNilssonPapageorgakis}, which we already employed in \cite{Me5}, with a focus on just an $U(1)$ part of $U(1) \times U(1)$ in that our pseudoscalars are neutral under $A_i^+ \equiv (A_i  + \hat{A}_i)$, and set $A_i^- \equiv (A_i  - \hat{A}_i)=0$.

Then, with the ansatz $\psi_{\hat{a}}^a = \frac{\delta_{\hat{a}}^a}{N} {\psi}$ for the singlet fermion and just one scalar with $Y = \frac{h(r)}{N} I_{N}$, where $h(r)$ is a scalar function on the boundary and $I_N$ is the unit $N\times N$ matrix \footnote{The conventions for $N$ coefficients of the fields and operators are according to \cite{Witten2}.}, from the main ABJM action \cite{Me4}, the scalar and fermion potentials vanish and so, we can write
\begin{equation}\label{eq56}
  \mathcal{L}_{3} = \mathcal{L}_{CS}^+ - \texttt{tr} \left(\bar{\psi} i \gamma^i \partial_i \psi \right) - \texttt{tr}\left(\partial_i Y^{\dagger} \partial^i Y \right) - \bar{W},
\end{equation}
where the CS Lagrangian and the last term (as a deformation), which comes in turn from the first line of (\ref{eq53a}) wrt (\ref{eq55}), read
\begin{equation}\label{eq56a}
  \mathcal{L}_{CS}^+ = \frac{i k}{4\pi}\ \varepsilon^{k ij}\ \texttt{tr} \left(A_i^+ \partial_j A_k^+ + \frac{2i}{3} A_i^+ A_j^+ A_k^+ \right), \quad \bar{W} \equiv \frac{\alpha}{N^2}\, \texttt{tr}\left((\partial_i Y^{\dagger} \partial^i Y) (\bar{\psi} \psi)\right),
\end{equation}
respectively. Then, to solve the EOM's for the scalar and fermion we use \footnote{For similar fermionic solutions see, for instance, \cite{Akdeniz1979}.}
\begin{equation}\label{eq56b}
  \begin{split}
     & h = b + \left( \frac{(x-x_0)_i (x-x_0)^i}{a\, \left(a^2 + (x-x_0)_i (x-x_0)^i \right)} \right)^{1/2}, \\
     & \psi= A\, \frac{\left(a + i (x-x_0)^i \gamma_i \right)}{\left[\left(a + i (x-x_0)^i \gamma_i \right) \left(a^{\dagger} - i (x-x_0)_i \gamma^{i\, \dagger} \right)\right]^{3/2}} \left(\begin{array}{c}   1  \\     0   \end{array}\right),
  \end{split}
\end{equation}
where $b$ is a boundary constant and $A$, as a normalization factor, comes from solving the coupled equations as $A = {a}/{\sqrt{3}}$ and note that $\alpha = \texttt{tr} (\psi \bar{\psi})^{-1}$. Still, from satisfying the $U(1)$ gauge equation, one could see that the magnetic charge because of $A^+$ vanishes \cite{Me5}. In addition, from this solution, one can confirm the state-operator correspondence in leading order according to (\ref{eq53a}).

As another aspect, it is interesting to use the approximate bulk solution (\ref{eq38CC}) in the latter case (with $\Delta_{\mp} = -2, +5$) and so (note that the perturbative solution does not break the SI in leading order),
\begin{equation}\label{eq57}
  \alpha = C_{9} \left( \frac{C_{10}}{\beta} \right)^{2/5},
\end{equation}
which acts as a triple-trace deformation (or deforming with a dimension-3 operator) \footnote{In general, the mixed boundary conditions lead to conformal field theories only if $f(\alpha) \sim \alpha^{3/\Delta_-}$ with $W=- \int d^3\vec{u}\, f(\alpha)$ or $\beta = f_0\, \alpha^{({3}/{\Delta_-})-1}$; and different values of $f_0$ corresponds to various points along the lines of marginal deformations. This statement is true for the leading order solutions (\ref{eq38CC}) as well.}. Then, the ansatz (\ref{eq56b}) solves the corresponding (pseudo)scalar and fermion equations with
\begin{equation}\label{eq58}
  A = - \frac{a^{\dagger}}{N} \frac{25}{10} \sqrt{30}\, 3^{3/10},
\end{equation}
where we have set $C_{9}=C_{10}=1$ for simplicity. One may also evaluate the finite correction for the main action based on the solution (\ref{eq56b}) from
\begin{equation}\label{eq59}
  W =\frac{1}{(3)^{2/5}} \int d^3\vec{u}\, \mathcal{O}_5^{3/5}, \quad \int_0^\infty \frac{r^2}{(a^2+r^2)^3}\, dr =\frac{\pi}{16\, a^3},
\end{equation}
with the boundary as a 3-sphere in infinity ($S_\infty^3$) and instanton localized at its center $\vec{u}_0=0$, and see that there is no dependence on $a$ because of the SI of the solution.

As a basic test, we see that
\begin{equation}\label{eq57b}
  \langle \mathcal{O}_{5} \rangle_{\alpha} \sim \frac{a^5}{\left(a^2 + (\vec{u}-\vec{u}_0)^2 \right)^5} \sim \beta= f(r)^{10/3},
\end{equation}
with $f(r)$ in (\ref{eq38CC}) and that, to the leading order, it corresponds to the bulk-to-bulk propagator and is $SO(4)$ invariant; see the discussion in section \ref{sec.5.1}.

\subsection{Duals For Massive Modes as Scalars} \label{sub.6.2}
Roughly speaking, it may be permissible to consider the modes as scalars in that $e_7$ and $J$ in the ansatz (\ref{eq11}) have the ingredients of the internal metric in (\ref{eq001b}) whose fluctuations produce the second and particularly third family of the scalars in the spectrum. Anyway, we take them as if they were scalars in this subsection. In this way, the operators $\Delta_+=4,5,6$ correspond to the reps $(n-2,2,0,0)$ of $0_3^+$ scalars with $n=2,4,6$ respectively. The corresponding operators emerge as the fourth descendant of $O^{4+k}$ in (\ref{eq53}) with $k=n-2$; see also Table 1 of \cite{EDHoker02}. As a result, a clear form for the operators reads
\begin{equation}\label{eq60}
     O_{\Delta_+=\frac{n}{2}+3} = \texttt{tr} \big(\Psi^{[I} \Psi^{J^\dagger} \Psi^{K} \Psi^{L^\dagger ]}\big)\, O^{n-2}.
\end{equation}
On the other hand, we recall from section \ref{sub.1.4} that there is no $H$-singlet from these modes under the branching $G\rightarrow H$ in the original theory; but after the swapping $\textbf{v} \leftrightarrow \textbf{s}$, which means exchanging the scalars and supercharges and is a way to account the breaking of SUSY as well, all above scalar modes include $H$-singlets. In this subsection, as an example, we make a solution based on this single-trace deformation for one of the operators. For $\Delta_+=4$, it might be considered as a double-trace deformation (see, for instance, \cite{Hartman} and references therein) of $\mathcal{O}_2^-=\psi \bar{\psi}$ of the conformally coupled pseudoscalar $m^2 R_{AdS}^2=-2$ already studied in \cite{Me5}. For $\Delta_+=5$, it may be considered as $\mathcal{O}_5=\mathcal{O}_4^- \mathcal{O}_1^+$, where the latter operator is indeed $\texttt{tr} (X^{[I} X^{J]})$ whose details were studied in \cite{I.N} and we have recently considered a special version of it in \cite{Me5} as $\mathcal{O}_1^+ =\frac{3}{4}\, \texttt{tr}(y \bar{y})$. Finally, for $\Delta_+=6$, we can consider the operator as $\mathcal{O}_6=\mathcal{O}_5^- \mathcal{O}_1^+ = (\mathcal{O}_2^-)^2(\mathcal{O}_1^+)^2$, which may be called a multi-trace combination of the single-trace operators; or may more precisely be considered as a double-trace deformation of the $\Delta_+=3$ operator taken in \cite{Me3}.

In fact, to find plain solutions in the latter cases, we first note to the swapping $\textbf{8}_v \leftrightarrow \textbf{8}_s$. As a result, the scalars set as $X^I \rightarrow (y^m, y, \bar{y})$ with $y = y^7 + i y^8$, $\bar{y}= y^\dagger$; and we focus on just the singlet one and a fermion. The remaining procedures are the same as those done in the previous subsection and we just comment on them briefly. Indeed, for the deformation with $\Delta_+=5$, the solutions for the scalar and fermion EOM's have the structures like those in (\ref{eq13a4}) and (\ref{eq56b}) with $a=0$, respectively. The same trend is applied to $\Delta_+=6$ deformation; and notice that wrt (\ref{eq38CC}) that suggests $\beta = f_0\, \alpha^{-2}$ with $f_0 = C_{10} C_{9}^{2}$, the resultant deformation turns again into that with a dimension-3 operator. We should also remind that for the solutions here, we have
\begin{equation}\label{eq61}
     \langle \mathcal{O}_{\Delta_+} \rangle_{\alpha} \sim {\left(a^2 + (\vec{u}-\vec{u}_0)^2\right)^{\Delta_+}} \sim \beta(\vec{u}),
\end{equation}
in the leading order of the perturbative solutions for which the SI is established.

It is also remarkable that the singlet mode $(n-2=0,0,0,0)$ of $0_2^+$ is for $\Delta_+=6$, which remains valid even after the swapping. The corresponding operator for this comes as the eighth descendant ($Q^8$) of $O^{4}$ in (\ref{eq53}) whose symbolic form is $\mathcal{O}_{\Delta_+=6} \approx (\partial X)^4$.

\section{Concluding Remarks}
In this study, by including a 7- and 4-form flux of 11D supergravity in ABJM background and from satisfying the equations and identities, we arrived at a second order NLPDE in Euclidean $AdS_4$, which could in turn be considered as a consistent truncation in that it led to a set of R-singlet (pseudo)scalars. Among the bulk modes, we focused on three massive ones and tried to gain solutions that were of course not exact. In fact, the bulk EOM (\ref{eq20b}) without the mass term and its solution (\ref{eq43a}) were $SO(4,1)$ invariant; but the mass term broke the symmetry into $SO(3,1)$ and as a result, the massless solution was valid only approximately for $m\, b_0\ll 1$. Then and after doing a bulk mass spectroscopy, we saw that the $SU(4) \times U(1)$ singlet modes might be realized in the spectrum if we considered exchanging the supercharge representation ($\textbf{8}_s$) with the scalar ($\textbf{8}_v$) and fermion ($\textbf{8}_c$) ones of $SO(8)$ in the original theory. The latter procedure guarantied the supersymmetry breaking scheme imposed by the bulk ansatz and solution as well.

For the massive modes from the bulk EOM, the boundary is changed by single-trace irrelevant operators that are not conserved currents. We proposed the standard forms for such operators, which were of course free ones because of the SIB and probable anomalous dimensions. Then, with respect to the AdS$_4$/CFT$_3$ correspondence rules, such as dual symmetry adjustments, we considered a truncated version of the ABJM boundary action with focusing on the $U(1)_{diag}$ of the quiver gauge group because of the parity breaking and a novel-Higgs mechanism valid there. After that, by taking suitable ansatzs, we made instanton solutions and confirmed the correspondence in leading order. We also showed that for a special perturbative bulk solution, we could make a mixed boundary condition associated with a marginal deformation preserving conformal symmetry as well-- note that with mixed boundary conditions, the parity symmetry is broken in general.

For more explanation, we read from the bulk ansatz that we have indeed added (anti)M5-branes to the Wick-rotated ABJM background, and the employed swappings were to achieve the desired singlet modes that broke all SUSY's as well. Besides, wrt the 7-form ansatz (\ref{eq11})--the term including $f_1$-- a domain-wall solution is possible. The situation is similar to Basu-Harvey equations \cite{BasuHarvey} that describe M2-branes ending on a M5-brane wrapped around a fuzzy $S^3/Z_k$, and in large $k$ limit they go to Nahm equations; see also \cite{Gomiz} and \cite{HanakiLin}. On the other hand, there are domain-wall flows that correspond to a thin-wall bubble of $AdS_-$ that expands exponentially within $AdS_+$; see \cite{Barbon1003}. The results agree with the picture that the boundary normalizable flows start from a local minimum of the SUGRA potential ($m^2 R_{AdS_+}^2 > 0$) and correspond to the bulk Coleman-de Luccia bounces that in turn break the conformal invariance spontaneously and so, come in continuous families associated with bulk translations of the $O(4)$ invariant solutions \footnote{Refer also to \cite{Maldacena010}, where it is shown that the field theories on $dS_3$ with irrelevant mass deformations and $O(3,1)$ invariant solutions are dual to vacuum decay processes and singularities in $AdS_4$.}.

\begin{wrapfigure}{r}{0.5\textwidth}
\centering
  \includegraphics[height=2.0in, width=2.20in, scale=1]{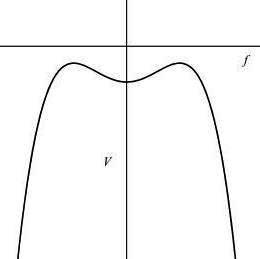}
  \caption{\textit{Scheme of the potential.}} \label{Fig1.}
\end{wrapfigure}

As a final point and related to the latter discussion, we return to the bulk potential and its stability issue. First, we remind that with the SIB, there was no exact solution with finite action and so argue that the vacua decay by approximate or constrained instantons. More precisely, we look at the bulk potential $V(f)=\frac{m^2}{2} f^2 - \frac{\lambda_4}{4} f^4$ with $m^2, \lambda_4 > 0$ (see Figure \ref{Fig1.}) that is arisen from the consistent truncation of the 11D SUGRA over $AdS_4 \times S^7/Z_k$ with the metric (\ref{eq001}) and the 4-form ansatz (\ref{eq18}) to the four dimensions of $AdS_4$. \\
The local minimum and maximums of the potential are in $f_0=0, f_{\pm}=\pm \frac{m}{\sqrt{\lambda_4}}$, respectively. This double-hump potential could be considered as an inverted double-well potential from which tunneling from $f_0$ to $f_{\pm}$ (or any arbitrary state on the slope) through both barriers is possible. A study of such potentials is already done in \cite{BumHoonLee013} with the solutions named as generalized Fubini instantons \footnote{We recall that the Fubini instantons \cite{Fubini1} represent tunneling from the top of a tachyonic potential to an arbitrary state instead of rolling down from a tachyonic potential including just a quadratic term.}. For a (pseudo)scalar sitting on a maximum (\emph{sphaleron} point), it is possible to run away to infinity or a domain-wall flow to $f_0=0$. In $AdS$, it is also possible to tunnel from the maximums to the states on the slope; and the solutions of the latter type are named as oscillating Fubini instantons \cite{BumHoonLee014}. Generally, with these unbounded potentials from below, the observables may prolong to infinity in finite times. Finally, it is interesting that our setup here agrees with an argument in \cite{OoguriVafa016} that any non-supersymmetric $AdS$ vacuum, which is supported by flux, must be unstable.

\end{document}